\documentclass{elsart}

\usepackage{graphics}
\usepackage{epsfig}

\usepackage{amssymb}

\begin{document}

\begin{frontmatter}



\title{Giant magnetic-field changes in radio-frequency absorption in
La$_{0.67}$Sr$_{0.33}$MnO$_3$ near the Curie temperature}

\author[ILTPE]{B. I. Belevtsev\corauthref{cor}},
\author[IRE]{A. Ya. Kirichenko},
\author[IRE]{N. T. Cherpak},
\author[IRE]{G. V. Golubnichaya},
\author[IRE]{I. G. Maximchuk},
\author[ILTPE]{E. Yu. Beliayev},
\author[ILTPE]{A. S. Panfilov},
\author[IPPAN]{J. Fink-Finowicki}

\address[ILTPE]{B. Verkin Institute for Low Temperature Physics \&
Engineering, Kharkov, 61103, Ukraine}

\address[IRE]{A. Usikov Institute for Radiophysics and Electronics,
National Academy of Sciences, Kharkov 61085, Ukraine}

\address[IPPAN]{Institute of Physics, Polish Academy of Sciences,
32/46 Al. Lotnikov, 02-668 Warsaw, Poland}
\corauth[cor]{Corresponding author. Email address:
belevtsev@ilt.kharkov.ua}

\begin{abstract}
The DC transport properties of and the radio-frequency (RF) wave
absorption (at 2.525 MHz) in a sample of
La$_{0.67}$Sr$_{0.33}$MnO$_{3}$ prepared by floating-zone method
are measured. The Curie temperature, $T_{\mathrm{c}}$, of the
sample is about 374~K. Giant temperature and magnetic-field
variations in RF absorption are found in the vicinity of
$T_{\mathrm{c}}$.  Relative change of the RF absorption in
magnetic field (magnetoabsorption) is about 67\% in field 2.1 kOe
and about 55\% in field 1 kOe. This giant magnetoabsorption effect
can be used to develop RF devices controlled by temperature and
low magnetic field. A weak temperature dependence of
magnetoabsorption for the sample studied in the range from room
temperature to about 350 K makes it especially attractive for
practical use. The RF study supplemented with transport,
magnetoresistive and magnetic measurements enables us to discuss
the optimal properties of manganite samples for observation of
giant magnetoabsorption in low field.
\end{abstract}

\end{frontmatter}

\section{Introduction}
In the last decade much interest is being shown in mixed-valence
manganites of the type La$_{1-x}$A$_x$MnO$_3$, where A is a
divalent alkaline-earth element like Ca, Sr, Ba
\cite{coey,tokura,dagotto}. In these compounds, a huge negative
magnetoresistance [so called, colossal magnetoresistance (CMR)
effect] has been found near the temperature, $T_{\mathrm{c}}$, of
paramagnetic-ferromagnetic transition. This effect offers
applications in advanced technology. The amplitude of CMR in
manganites depends essentially on magnitude of applied magnetic
field and the value of Curie temperature $T_c$ \cite{coey,tokura}.
For example, in optimally doped La$_{1-x}$A$_x$MnO$_3$ with
$x\approx 0.3$, high-field (60--100 kOe) magnetoresistance (MR),
defined as $-\Delta R(H)/R(H)$, can be about 5 at $T\simeq T_c$ in
Ca manganites with $T_c \simeq 250$~K; whereas in Sr manganites
with $T_c$ about 370 K  the MR is by the order of magnitude
smaller \cite{coey,zhang}.
\par
In spite of enormous efforts, a clear understanding of CMR is not
yet available, though some numerical simulations can describe
experimental data with an acceptable accuracy \cite{coey,dagotto}.
Beside this, the requirement of high magnetic field hampers a
possible application of the CMR manganites to practical use. It is
desirable, therefore, for application purposes to look for some
other ``non-magnetic'' properties of manganites which are
dependent, however, on their magnetic properties to the extent
that it can ensure  a sensitivity to rather low magnetic field. A
well known property of this type is the high-frequency complex
resistance (impedance) which can show a high sensitivity to low
magnetic field in soft ferromagnetic (FM) conductors. Generally
this phenomenon is called magnetoimpedance. In the case that a
considerable relative change (several tens of percent) in the
impedance takes place in low external magnetic field this
phenomenon is referred to as giant magnetoimpedance \cite{knobel}.
\par
The impedance of a sample is determined by its resistance and
reactance both of which are influenced by absorption of applied
electromagnetic waves. It is known that in FM metals this
absorption is influenced not only by the conductivity, but by the
magnetic permeability, $\mu_{r}$, of absorbing media as well. The
skin depth in FM conductor is given by
\begin{equation}
\delta=\sqrt{\frac{2\rho}{\omega \mu_{0}\mu_{\mathrm{r}}}},
\end{equation}
where $\rho$ is the resistivity, $\omega=2\pi\nu$ the wave angular
frequency, and $\mu_{0}$ the magnetic permeability of vacuum. The
application of an external magnetic field leads to a considerable
decrease in the permeability, causing increase in the skin depth,
and hence a giant change in the high-frequency wave absorption.
This effect is, however, fairly large only for optimal frequency
range. On the one hand, the frequency must be high enough to
assure that $\delta$ is smaller than the sample dimensions
\cite{knobel}. On the other hand, the magnitude of
$\mu_{\mathrm{r}}$ in zero magnetic field should be high enough to
produce a giant magnetoabsorption (or magnetoimpedance) effect.
Really, it is known that the value of $\mu_{\mathrm{r}}$ in FM
metals decreases with increasing frequency going close to unity
for $\nu \geq 100$~MHz \cite{vons}. At high enough frequency,
therefore, a FM metal behaves like a non-magnetic one. For this
reason, the optimal frequency for observation of giant changes in
high-frequency absorption should be of the order of 1 MHz, that
is, in the radio-frequency (RF) range. For example, a giant RF
magnetoabsorption is found in cobaltite
La$_{0.5}$Sr$_{0.5}$CoO$_{3}$ \cite{boris} which is a FM
perovskite-like oxide related to manganites but with a far smaller
MR. The relative change in magnetoabsorption in this compound was
found to be about 38 \% near the Curie temperature (about 250 K)
in a rather weak magnetic field 2.1 kOe at frequency 1.33 MHz
\cite{boris}.
\par
In this article, we report a study of RF magnetoabsorption effect
in a bulk sample of La$_{1-x}$Sr$_{x}$MnO$_{3}$ ($x=0.33$)
prepared by floating-zone method. For this Sr concentration,
according to the known phase diagram \cite{tokura}, manganite
samples are in metallic state both below and above $T_c$. The
$T_c$ was about 374 K in the sample studied.  The previous RF
studies of magnetoimpedance in Sr manganites (with about the same
composition, $x=0.3$ and 0.35)  were done \cite{hu}  only at room
temperature, that is rather far below $T_c$. A rather minute
magnetoimpedance effect (a few percent) is found in these samples.
Temperature dependence of the complex permeability in
La$_{0.7}$Sr$_{0.3}$MnO$_{3}$ and the effect of crystal-lattice
disorder on this were studied in Refs. \cite{wang1,wang2}. All
those studies were done on polycrystalline ceramic samples.  In
this study, RF (2.525 MHz) absorption in
La$_{0.67}$Sr$_{0.33}$MnO$_{3}$ sample was studied in rather large
temperature range (from room temperature up to about 400 K), which
includes $T_c$. In contrast to Refs. \cite{hu}, we found a really
giant magnetoabsorption below $T_{\mathrm{c}}$ (including room
temperature) with maximum amplitude about 67 \% in low magnetic
field 2.1 kOe. The effect remains to be giant in smaller fields as
well. For example, in field 1 kOe the magnetoabsorption is about
55~\%. This makes it attractive for practical use. RF study was
supplemented with transport, magnetoresistive and magnetic
measurements of the sample which make it possible to discuss the
optimal properties of manganite samples for observation of giant
magnetoabsorption in low field.

\section{Experimental}
A crystal of nominal composition La$_{0.67}$Sr$_{0.33}$MnO$_3$ was
grown by the floating zone method at the Institute of Physics,
Warsaw. The appropriate amounts of starting materials La$_2$O$_3$,
SrCO$_3$ and MnO$_2$ were calcined at temperature 1000$^{\circ}$C,
then mixed, compacted into pellets and sintered at
1400$^{\circ}$C. After that the pellets were milled, and resulting
powder was pressed to form the feed rod with a diameter of 8 mm
and a length of 90 mm. This rod was further sintered at
1470$^{\circ}$C during 12 h in air. An optical furnace (type
URN-2-3Pm made by Moscow Power Engineering Institute) with two
ellipsoidal mirrors and 2500 W xenon lamp as the heat source was
applied for crystallization. The feed rod and the growing crystal
were rotated in opposite directions to make heating uniform and to
force convection in the melting zone. The growth rate was 1 mm/h.
An additional afterheater was used for lowering the temperature
gradients in growing crystal. It is known \cite{fink} that
manganite crystals produced by the described technique are like
single crystals\footnote{For example, our X-ray diffraction study
of related manganite crystal, La$_{0.67}$Ca$_{0.33}$MnO$_3$,
prepared in Institute of Physics, Warsaw, in the same
floating-zone equipment has revealed that this crystal is very
close to a single-crystal state although with twins which are
inevitable arisen in manganites grown by floating-zone method.},
and, in this respect, they have far better crystal quality and far
less porosity than the samples prepared by a solid-state reaction
technique. On the other hand, the crystals have twin domain
structure. X-ray powder diffraction pattern has testified a
single-phase state of the sample.
\par
The magnetization $M$ of the sample studied was measured in a
Faraday-type magnetometer. The DC resistance, as a function of
temperature and magnetic field $H$ (up to 16 kOe), was measured
using a standard four-point probe technique. The RF technique
employed is essentially the same as that used by some of us for
ceramic cobaltites \cite{boris}. The sample for RF measurements
(with diameter about 8 mm and length about 10 mm) was placed in an
induction coil (9.7 mm in diameter and with a height about 22 mm).
This coil is a part of an {\it LC} tank circuit. The magnitude of
RF magnetic field in the induction coil was not more than 0.1 Oe.
The measurements of the quality factor, $Q$, of the {\it LC} tank
with and without the sample inside the coil were taken. The
measurements of the $Q$-factor depending on temperature or
magnetic field  were taken at fixed frequency $\nu =2.525$~MHz.
The available cryostat with electromagnet makes it possible to
measure the $Q$-factor in DC magnetic fields up to 2.5 kOe. The DC
and RF fields were mutually perpendicular. The temperature and
magnetic-field dependences of the $Q$-factor were recorded mainly
on heating after the sample had been cooled down in zero field.
\par
Consider shortly the physical properties which can be derived from
the $Q$-factor measurements of the RF {\it LC} tank employed. The
sample behavior in an electromagnetic field can be described,
using complex permittivity $\epsilon_{\mathrm{c}} =
\epsilon^{'}-i\epsilon^{''} = \epsilon^{'}\left[ 1-i\tan
(\delta_{\epsilon})\right]$ (where
$\epsilon^{'}=\epsilon_{r}\epsilon_{0}$, $\tan (\delta_{\epsilon})
= \sigma_{\mathrm{rf}}/(\epsilon^{'} \omega)$ is the dielectric
loss tangent, $\sigma_{\mathrm{rf}}$ is the RF conductivity of the
sample) and complex magnetic permeability
$\mu_{\mathrm{c}}=\mu^{'} - i\mu^{''} = \mu^{'}\left[
1-i\tan(\delta_{\mu})\right]$, where
$\mu^{'}=\mu_{\mathrm{r}}\mu_{0}$ and $\tan(\delta_{\mu})=
\mu^{''}/\mu^{'}$. For a plane electromagnetic wave, the complex
resistance (impedance) of the sample is given by relation
\begin{equation}
Z_{\mathrm{c}}=
\sqrt{\frac{\mu_{\mathrm{c}}}{\epsilon_{\mathrm{c}}}} =
R_{\mathrm{c}}+iX_{\mathrm{c}},
\end{equation}
where resistance $R_{\mathrm{c}}$ defines the energy loss in the
sample, and $X_{\mathrm{c}}$ is the sample reactance.
\par
Once the sample with the wave resistance $Z_{\mathrm{c}}$ is
inserted into the coil of the RF circuit, arranged from
resistance, inductance and capacitance elements $R_0$, $L_0$ and
$C_0$, a complex resistance is added to the circuit, given by
$Z_{\mathrm{sp}}=kZ_{\mathrm{c}}$ where $k$ is a geometrical
factor determined by the sample and coil dimensions, which was
about 0.35 in this study. The $Q$-factor of the circuit without
the sample is expressed as $Q_0 = \omega L_{0}/R_0$. It is often
more convenient to use the  damping factor (or decrement)
$d=Q^{-1}$. The decrement of the circuit without the sample is,
therefore, $d_0=R_{0}/\omega L_{0}$. On inserting the sample into
the coil the circuit decrement changes to $
d=(R_{0}+kR_{\mathrm{c}})/[\omega(L_{0}+kL_{\mathrm{c}})]. $ In
the case that the sample causes merely small changes in the
circuit inductance ($kL_{\mathrm{c}}\ll L_{0}$) and the circuit
quality without the sample is fairly high ($\omega L_{0}/R_{0} \gg
1$), the expression for the difference $d-d_0=P_{\mathrm{A}}$ can
be written as
\begin{equation}
P_{\mathrm{A}}=k\frac{R_{\mathrm{c}}}{\omega L_{0}}
=k\frac{R_{\mathrm{c}}} {R_{0}Q_{0}}. \label{eq}
\end{equation}
It is seen from it that $P_{\mathrm{A}}$ is determined by the loss
in the sample. For the further analysis of the situation it is
necessary to make suggestions about the amplitudes of dielectric
and magnetic loss tangents. Taking $\tan (\delta_{\epsilon})\gg 1$
and $\tan(\delta_{\mu})\ll 1$, the Eq. (\ref{eq}) can be presented
in the form
\begin{equation}
P_{\mathrm{A}}=k_0 \left
(1+\frac{1}{2}\frac{\mu^{''}}{\mu^{'}}\right ) R_s, \label{p2}
\end{equation}
where $k_{0} =k/(\omega L_{0})$,

\begin{equation}
R_s=(\omega\mu^{'}/2\sigma_{\mathrm{rf}})^{1/2}=
(\omega\mu^{'}\rho_{\mathrm{rf}}/2)^{1/2} \label{rs}
\end{equation}
is the surface resistance, $\rho_{\mathrm{rf}}$ is the RF
resistivity. According to Refs. \cite{wang1,wang2},
$\tan(\delta_{\mu})$ is close to unity in polycrystalline samples
of La$_{0.7}$Sr$_{0.3}$MnO$_3$ at the frequency used in this
study. The same can be expected for the sample studied as well.
In the case $\tan (\delta_{\epsilon})\gg 1$ and arbitrary
magnitude of $\tan(\delta_{\mu})$, the expression for $P_A$ takes
the form
\begin{equation}
P_{\mathrm{A}}=k_0 \left [\frac{\cos \left (
\delta_{\mu}/2)+\sin(\delta_{\mu}/2 \right
)}{\sqrt{\cos(\delta_{\mu})}} \right ] R_s. \label{p3}
\end{equation}
It is seen that in any case the loss behavior is determined mainly
by that of the surface resistance $R_s$.

\section{Results and discussion}
The temperature dependence of DC magnetization, $M$, of the sample
studied is shown in Fig.~1. The value of $T_{\mathrm{c}}\approx
374$~K can be found if $T_{\mathrm{c}}$ is defined as the
temperature of the inflection point in the $M(T)$ curve. The value
of $T_c$ and general behavior of $M(T)$ agree well with previous
studies of La$_{1-x}$Sr$_{x}$MnO$_3$ samples with $x=0.3-0.35$
\cite{zhang,urushibara,mukhin}. The inset in Fig.~1 shows the
magnetic-field dependence of the magnetization at $T=290$~K. It is
seen that the magnetization is close to saturation already at
rather low field about 1 kOe. Taking into account a saturation
value of $M$ at higher fields, we have obtained the magnetic
moment per formula unit at $T=290$~K to be equal to $\mu_{fu}=
2.95$~$\mu_B$, where $\mu_B$ is the Bohr magneton. Extrapolating
$M(T)$ to low temperature limit ($T\rightarrow 0$) for higher
fields enables us to obtain $\mu_{fu}(0)=3.8\pm 0.1$~$\mu_B$. For
nominal composition La$_{1-x}$Sr$_x$MnO$_3$ ($x=0.33$) of the
sample studied, taking into account that spin of Mn$^{+3}$ is
$S=2$ and that of the Mn$^{+4}$ is $S=3/2$, $\mu_{fu}$ should be
equal to $(4-x)$~$\mu_B$, that is to 3.67 $\mu_B$. This value
agrees rather well with the  estimated value, indicated above.
\par
The temperature dependence of DC resistivity, $\rho (T)$, is shown
in Fig.~2. The $\rho (T)$ behavior is metallic
($\mathrm{d}\rho/\mathrm{d}T > 0$) over the whole temperature
range investigated, below and above $T_{\mathrm{c}}$. The $\rho$
magnitude decreases enormously (by a factor of 40) when going from
400 K to 77 K (Fig.2). The $\rho$ values in low temperature range
are fairly low (about $10^{-4}$~$\Omega$~cm). These are even less
than those found in Ref. \cite{urushibara} in
La$_{1-x}$Sr$_{x}$MnO$_3$ ($x=0.3$) samples prepared (the same as
in this study) by floating zone method, but are equal to those
found in single-crystal samples of the same composition
\cite{mukhin}. This is an evidence of excellent crystal perfection
of the sample studied. The $\rho(T)$ dependence exhibits a change
of slope at $T\approx T_{\mathrm{c}}$ (see inset in Fig.~2) due to
a contribution from the electron scattering on the spin disorder
(in addition to the usual contributions from crystal lattice
defects and electron-phonon scattering) \cite{vons}. This
``magnetic'' contribution to the resistivity, $\Delta \rho_{m}$,
depends on the magnetization. The external magnetic field enhances
the spin order (that is, the magnetization), which leads to a
decrease in the resistivity, that is to negative MR.
\par
The magnetic-field dependences of resistance of the sample studied
are shown in Fig.~3. The temperature dependences of the MR at
different amplitudes of $H$ are presented in Fig.~4. Both these
figures show that the MR depends strongly on temperature and is
maximal near $T_c$. It goes down rather steeply for temperature
deviating to either side from $T_{\mathrm{c}}$. This temperature
behavior of the MR is expected for FM manganites of fairly good
crystal perfection.  Indeed, the MR amplitude is determined by the
ability of an external magnetic field to increase the
magnetization. It is obvious that at low temperature ($T\ll
T_{\mathrm{c}}$), when nearly all spins are already aligned by the
exchange interaction, this ability is minimal. For increasing
temperature and, especially, at temperature close to
$T_{\mathrm{c}}$, the magnetic order becomes weaker (the
magnetization goes down) due to thermal fluctuations. In this case
the possibility to strengthen the magnetic order with an external
magnetic field increases profoundly. This is the reason for
maximal MR magnitude near $T_{\mathrm{c}}$. Above
$T_{\mathrm{c}}$, the spin arrangement becomes essentially random,
the magnetization is zero, and, therefore, the MR is close to zero
as well.
\par
Let us now take a look at behavior of the RF loss $P_{\mathrm{A}}$
in the sample studied. We begin with the temperature behavior
$P_{\mathrm{A}}(T)$ in zero field (Fig.~5). It is seen that a
sharp (three-fold) increase in $P_{\mathrm{A}}$ takes place when
temperature goes below $T_{\mathrm{c}}$. This is determined by
correlation between the temperature behavior of the loss and that
of the permeability $\mu_{\mathrm{r}}$ \cite{knobel,boris}. It is
clear that far enough above $T_{\mathrm{c}}$ the relations
$\mu_{\mathrm{r}}=1$ and $\mu^{''}=0$ are true. In this case [see
Eqs.~(\ref{p2}), (\ref{rs}) and (\ref{p3})] the loss depends only
on the conductivity, that is $P_A$ should decrease with
temperature decreasing.  In mixed-valence manganites, the FM
fluctuations (or developing of small FM regions in the
paramagnetic matrix) can be evident even fairly far above
$T_{\mathrm{c}}$ for extrinsic and intrinsic sources of magnetic
inhomogeneity \cite{dagotto,boris2}. In that event, as
$T_{\mathrm{c}}$ is approached from above, the slight increase in
the permeability can occur even in the range above
$T_{\mathrm{c}}$. This will cause a corresponding increase in the
loss. The drastic increase in $P_{\mathrm{A}}$ occurs, however,
only quite near or somewhat below $T_{\mathrm{c}}$ when transition
of the most of the sample volume to the FM state takes place
inducing a sharp increase in $\mu^{'}$.
\par
A decrease in temperature below $T_{\mathrm{c}}$ down to room
temperature leads only to very weak changes in $P_A$ (Fig. 5). In
the case that the loss is determined mainly by the surface
resistance $R_s$ [see Eq. (\ref{rs})], the weak temperature
dependence of $P_A$ means that the magnitude of
$\mu^{'}\rho_{\mathrm{rf}}$ is nearly constant in this temperature
range. Since $\rho$ decreases rather considerably with temperature
decreasing in this temperature range, this implies that the
permeability should concurrently increase to compensate the $\rho$
decreasing. This behavior is possible basically  in Sr manganites.
For example, according to Ref. \cite{wang1}, the real part of the
permeability, $\mu_r$, in polycrystalline samples of
La$_{0.7}$Sr$_{0.3}$MnO$_3$ increases considerably with
temperature decreasing below $T_{\mathrm{c}}$ at frequency 10 MHz
(although it decreases at $\nu = 1$~MHz and for lower
frequencies).
\par
It is seen that the RF loss is far more sensitive to strengthening
of the spin order at the transition to the FM state than the DC
conductivity. Really, $\rho (T)$ shows only a change in the slope
at $T_{\mathrm{c}}$ that means the steeper decrease in resistivity
with temperature decreasing below  $T<T_{\mathrm{c}}$ (Fig.~2). In
contrast to this, the RF loss increases sharply (by a factor of
three) in the vicinity of $T_{\mathrm{c}}$ in zero field (Fig.~5).
It is the change in the permeability that causes the large
variations in the loss near $T_{\mathrm{c}}$.
\par
From measurements of the loss, $P_{\mathrm{A}}$, at different
magnetic fields we have obtained the values of the
magnetoabsorption (MA), which has been defined as
$[P_{\mathrm{A}}(H)-P_{\mathrm{A}}(0)]/P_{\mathrm{A}}(0)= \Delta
P_{\mathrm{A}}(H)/P_{\mathrm{A}}(0)$. Under suggestion, that the
loss is determined primarily by the surface resistance
$R_{\mathrm{s}}$, and taking into account Eq.~(\ref{rs}), a
simplified expression for MA can be written as
\begin{equation}
\Delta P_{\mathrm{A}}(H)/P_{\mathrm{A}}(0)\approx
\frac{R_{s}(H)}{R_{s}(0)}-1=
\sqrt{\frac{\mu^{'}(H)\rho_{\mathrm{rf}}(H)}
{\mu^{'}(0)\rho_{\mathrm{rf}}( 0)}}-1. \label{ph}
\end{equation}
This shows the correlation between the $P_{\mathrm{A}}(T)$ and
$\mu^{'}(T)$. It is clear from Eqs.~(\ref{rs}) and (\ref{ph}) that
owing to decrease in $\rho_{\mathrm{rf}}$ and $\mu^{'}$ in an
applied magnetic field, the MA should be negative, as it was
really found in this study. We discovered an extremely high
sensitivity of the RF loss to applied magnetic field (Figs.~5, 6
and 7). It is remarkable that the strong magnetoabsorption effect
shows itself in rather low magnetic fields.  Not far from
$T_{\mathrm{c}}$ the MA is about 67\% in field $H=2.1$~kOe
(Fig.~6). By contrast, the DC MR is of only about 4 \% near
$T_{\mathrm{c}}$ in the above-mentioned field (Fig.~3). The MA
remains to be very high even for lower field. For example, in
field 1 kOe the MA is about  55 \%.
\par
A fairly high magnitude of the permeability at zero field is a
necessary condition to ensure giant MA (or magnetoimpedance)
effect. In FM oxides this can be achieved only under the condition
of their high crystal perfection. It is known, for example, that
the permeabilty of ferrites decreases dramatically with increasing
in porosity and decreasing in grain size \cite{smit}.  The same is
true for manganites. According to Ref. \cite{wang2}, the
permeability of ceramic La$_{0.7}$Sr$_{0.3}$MnO$_3$ samples
reduces severely with grain size decreasing. In this case high MA
magnitude can not be expected. For this reason, obviously, rather
low (about 3 \%) magnetoimpedance was found in ceramic samples of
La$_{0.67}$Sr$_{0.33}$MnO$_3$ prepared by solid-state reaction
method. By contrast, in the sample studied prepared by
floating-zone method, which has much better crystal perfection and
negligible porosity, the MA is found to be really giant (close to
70 \%, as was indicated above).
\par
In conclusion, we found the giant negative RF magnetoabsorption
effect in bulk sample of La$_{0.67}$Sr$_{0.33}$MnO$_{3}$ under low
DC magnetic fields. The effect is attributed primarily to the
decrease in the magnetic permeability in applied magnetic field.
Such low-field magnetoabsorption may be of potential application
in RF devices controlled by low magnetic field in room temperature
range. Especially attractive for some purposes can be a weak
temperature dependence of the giant magnetoabsorption in the range
from room temperature to about 350 K found in the sample studied.

\newpage
\centerline{\bf{Figure captions}} \vspace{15pt} Figure 1.
Temperature dependence of the DC magnetization of the sample
La$_{0.67}$Sr$_{0.33}$MnO$_3$ in magnetic field $H=5.4$~kOe. The
dependence was recorded with temperature increasing after the
sample was cooled down to liquid nitrogen temperature, $T\approx
77.3$~K, in a field close to zero. The inset shows the magnetic
field dependence of the magnetization at $T=290$~K. \vspace{15pt}

Figure 2. Temperature dependence of the DC resistivity and its
derivative $\mathrm{d}\rho/\mathrm{d}T$ (inset) of the sample
studied. \vspace{15pt}

Figure 3. Magnetoresistance curves of the sample studied for
different temperatures above $T_c$ (a) and below $T_c$ (b).
\vspace{15pt}

Figure 4. Temperature dependence of magnetoresistance at
$H=16$~kOe, 10 kOe and 3 kOe. The solid lines present a B-spline
fitting. \vspace{15pt}

Figure 5. Temperature dependences of RF absorption ($\nu
=2.525$~MHz) in the sample studied  in zero field and in the field
$H=2.1$~kOe. $P_{\mathrm{A}}= d-d_0$, where $d$ and $d_0$ are the
decrements of the used LC circuit with the sample in an inductance
coil and without the sample in it, respectively. \vspace{15pt}

Figure 6. Temperature dependence of the RF magnetoabsorption in
the sample studied in field $H=2.1$~kOe. \vspace{15pt}

Figure 7. Magnetic-field dependences of the RF magnetoabsorption
in the sample studied at room temperature. The curves show a
hysteretic behavior. The arrows indicate directions of the field
sweeps. The filled triangles correspond to $P_A(H)$ behavior for
initial increase in magnetic field. The solid lines are guides to
the eye. \vspace{15pt}

\newpage
\begin{figure}[htb]
\centering\includegraphics[width=95mm]{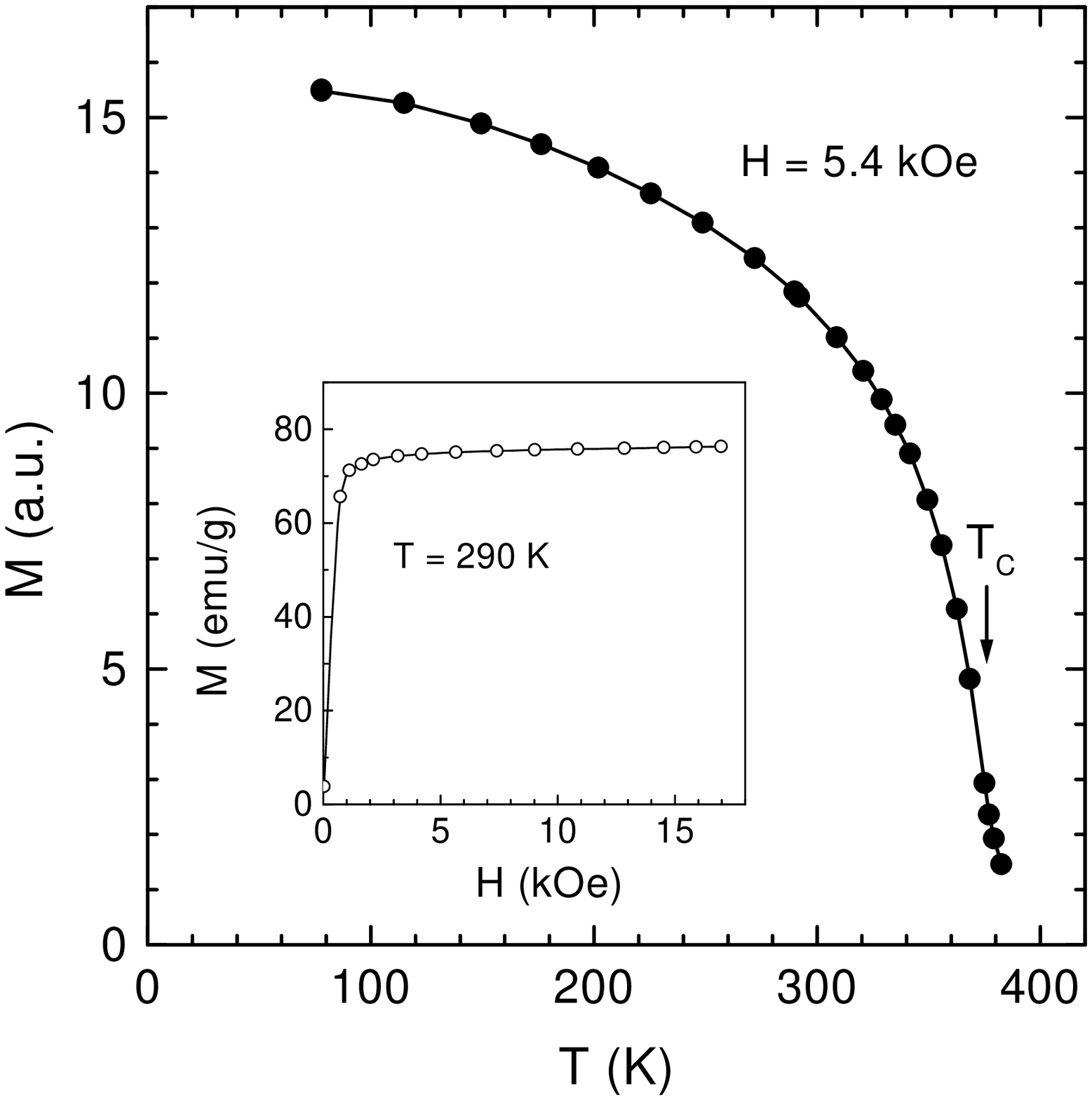}
\centerline{Figure 1 to paper Belevtsev et al.}
\end{figure}

\begin{figure}[htb]
\centering\includegraphics[width=100mm]{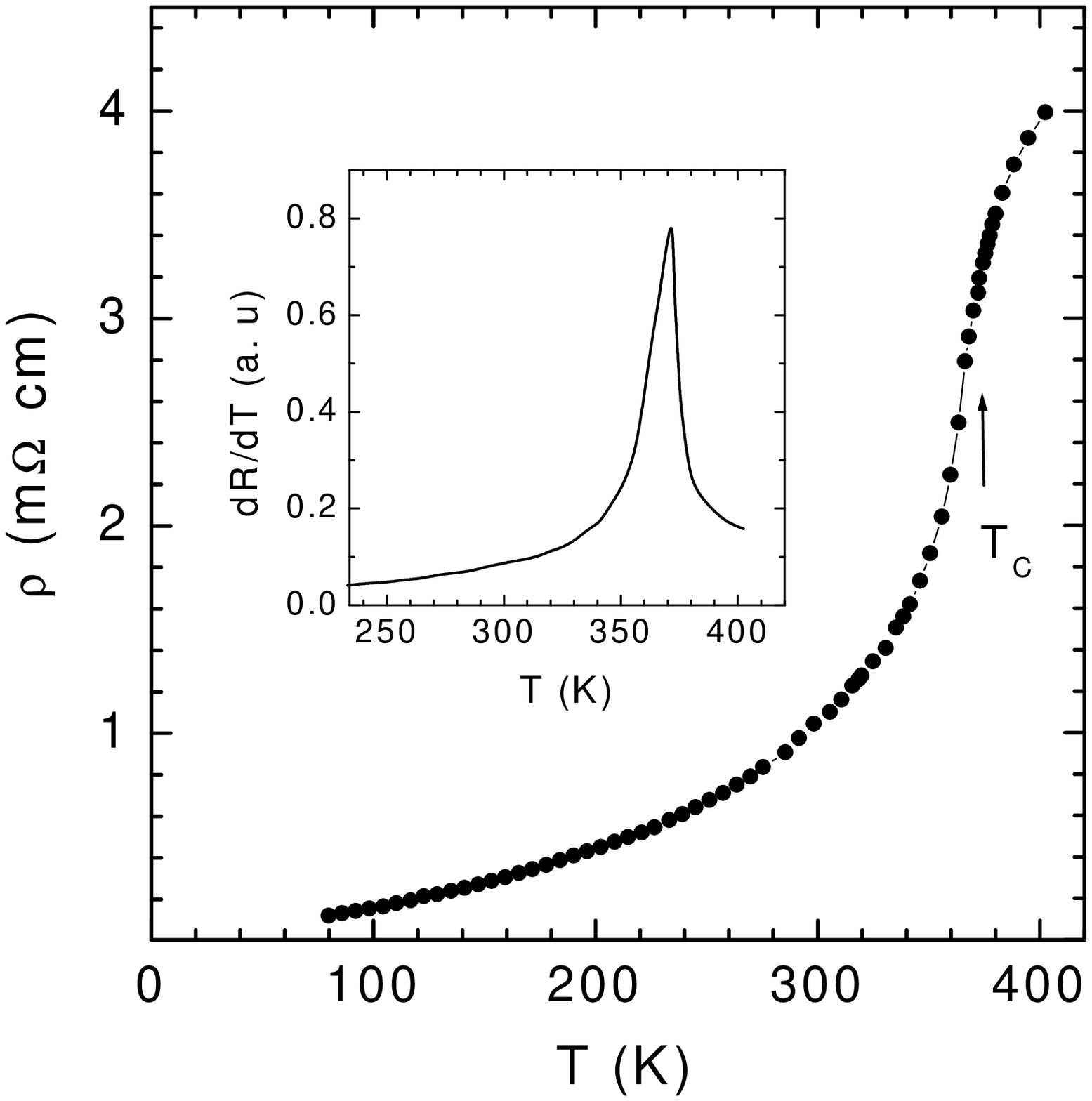}
\centerline{Figure 2 to paper Belevtsev et al.}
\end{figure}
\newpage
\begin{figure}[htb]
\centering\includegraphics[width=120mm]{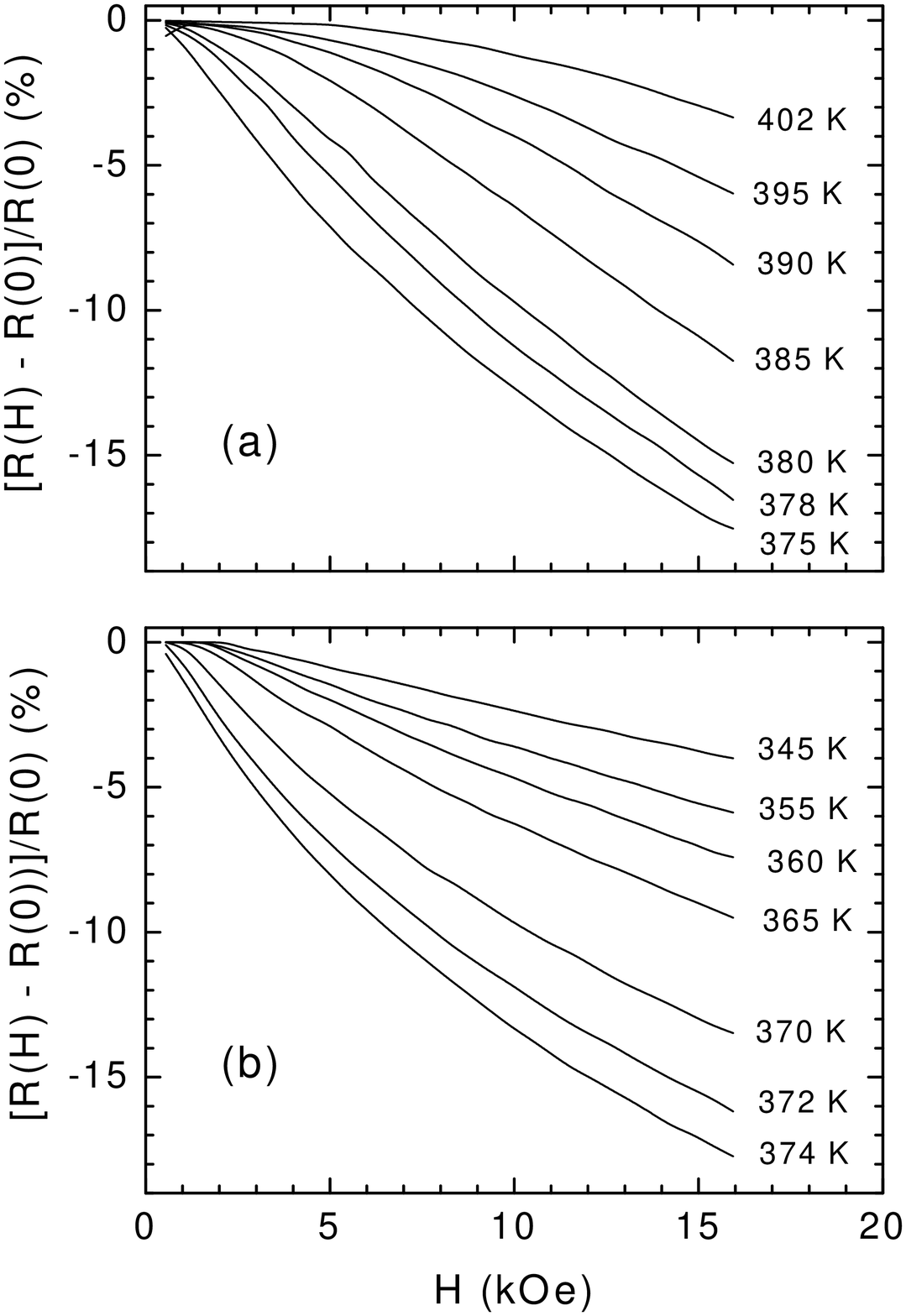}\vskip 25pt
\centerline{Figure 3 to paper Belevtsev et al.}
\end{figure}

\newpage
\begin{figure}[htb]
\centering\includegraphics[width=95mm]{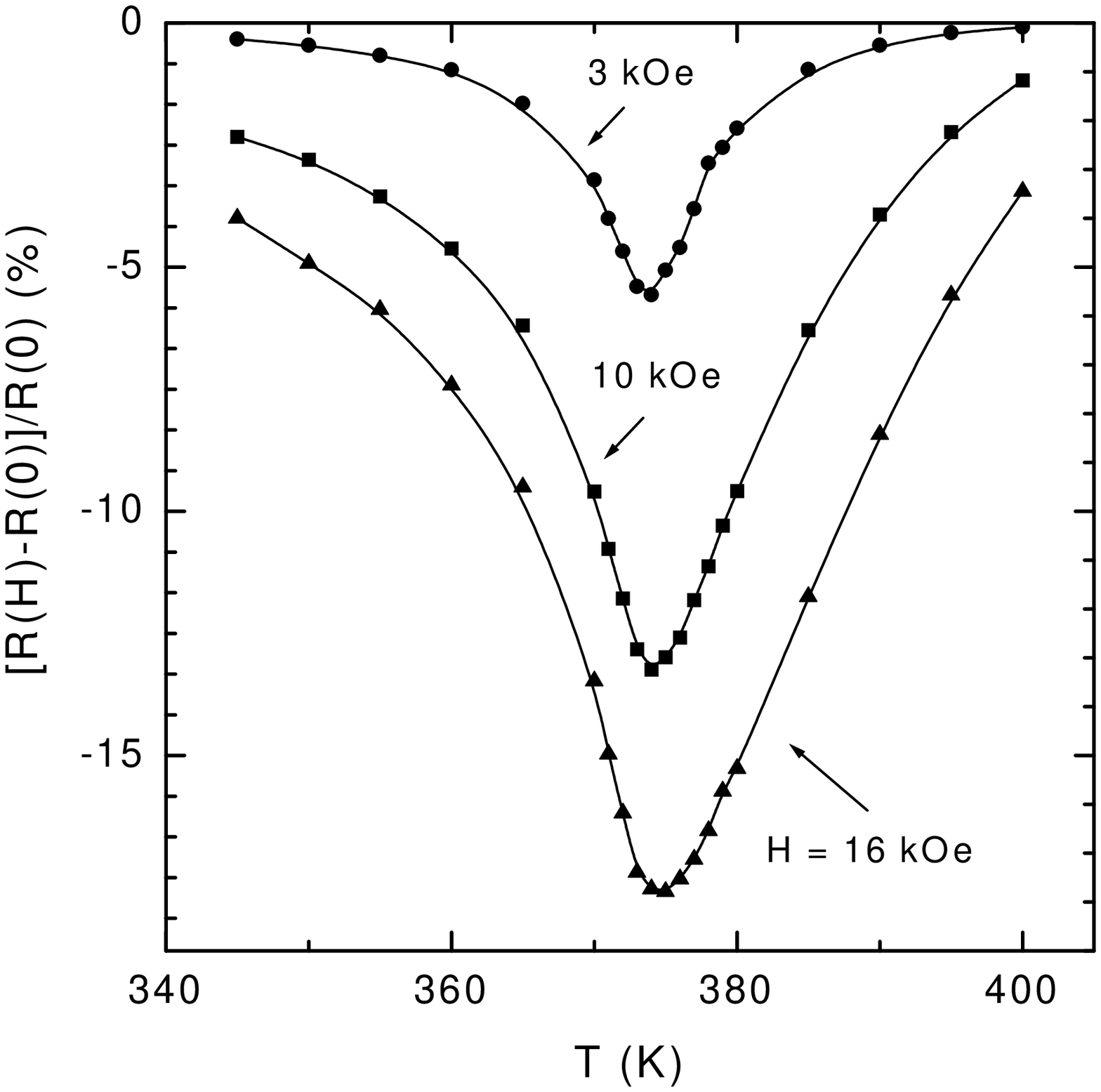}
\centerline{Figure 4 to paper Belevtsev et al.}
\end{figure}

\begin{figure}[htb]
\centering\includegraphics[width=95mm]{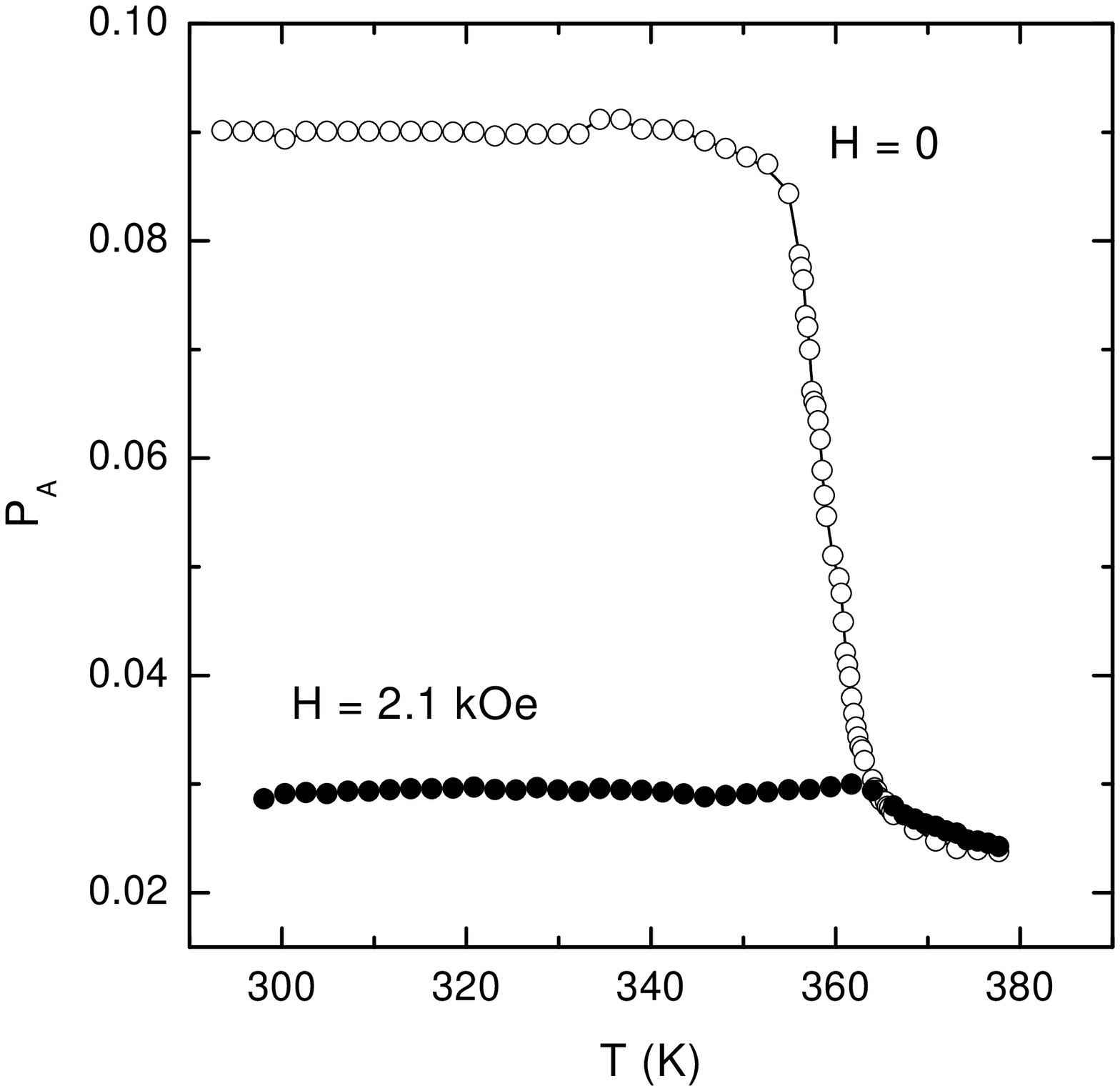}
\centerline{Figure 5 to paper Belevtsev et al.}
\end{figure}

\newpage
\begin{figure}[htb]
\centering\includegraphics[width=95mm]{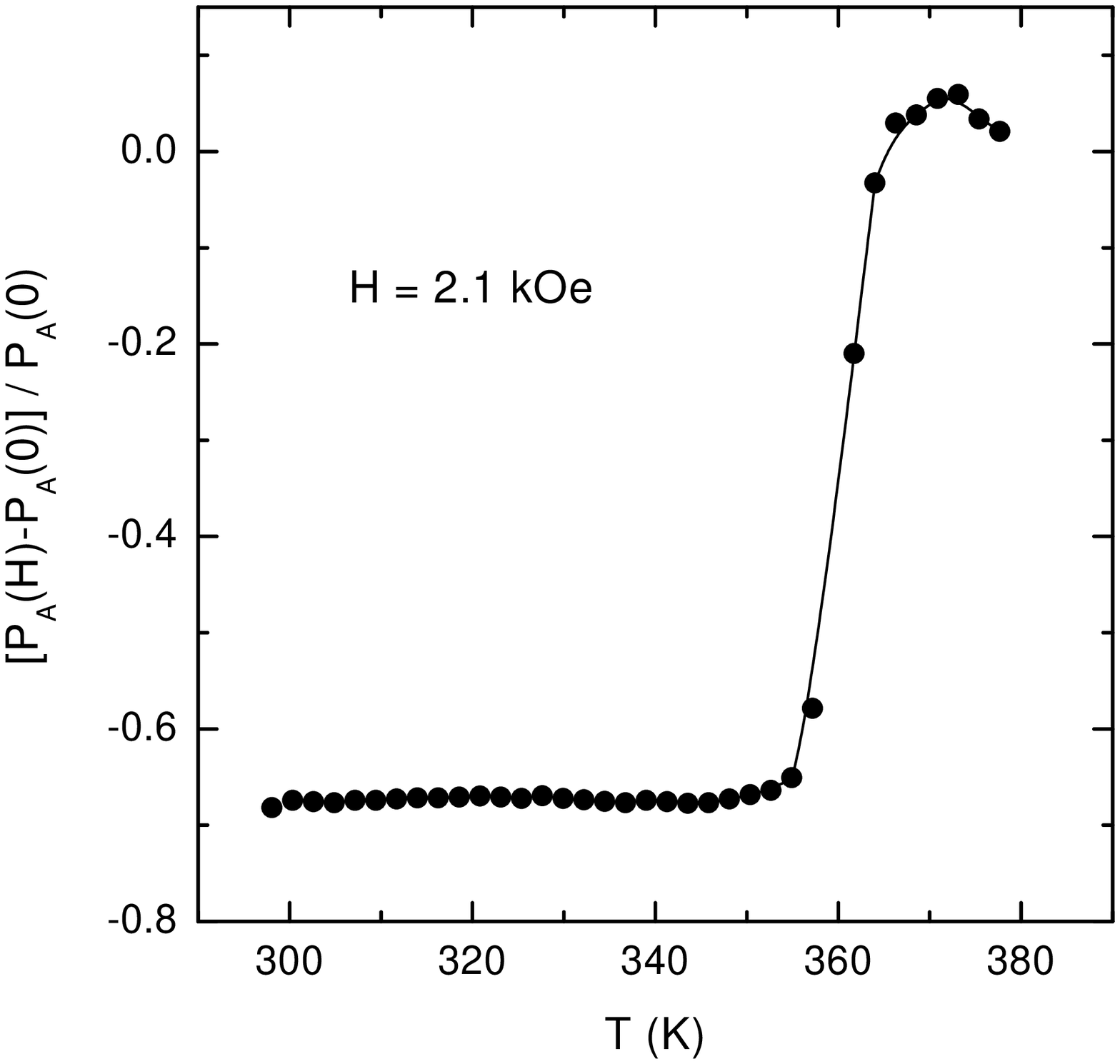}
\centerline{Figure 6 to paper Belevtsev et al.}
\end{figure}

\begin{figure}[htb]
\centering\includegraphics[width=95mm]{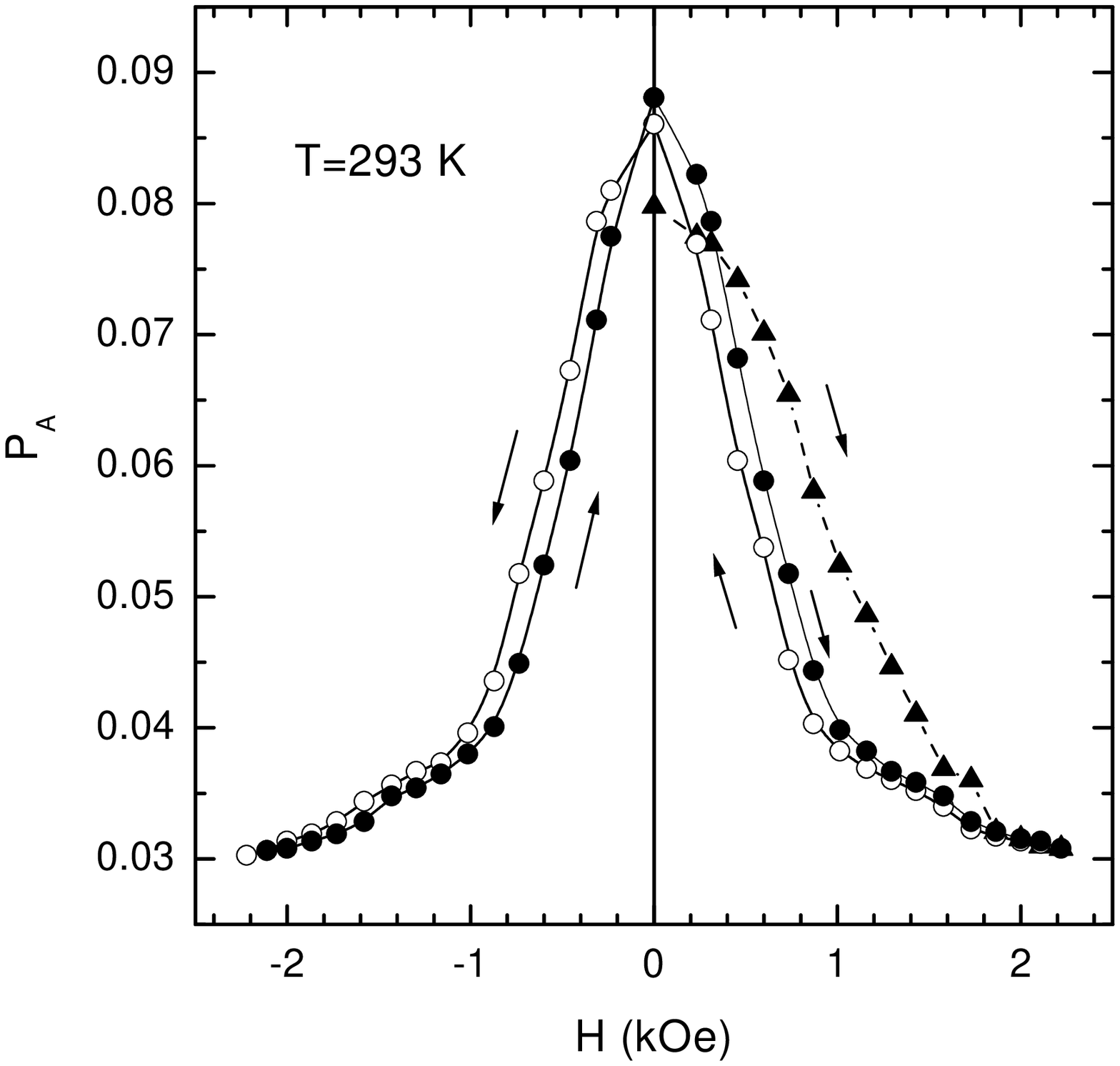}
\centerline{Figure 7 to paper Belevtsev et al.}
\end{figure}


\begin{thebibliography}{00}
\bibitem{coey}J. M. D. Coey, M. Viret, and S. von Molnar, Adv. Phys. 48 (1999)
167.

\bibitem{tokura}{\it Colossal magnetoresistive oxides}, ed. by Y.
Tokura, Gordon and Breach Science Publ., 2000.

\bibitem{dagotto}E. Dagotto, T. Hotta, and A. Moreo, Phys. Rep. 344 (2001) 1.

\bibitem{zhang}Ning Zhang, Fang Wang, Wei Zhong, and Weiping Ding,
J. Phys.: Condens Matter 11 (1999) 2625.

\bibitem{knobel}M. Knobel and K. R. Pirota, J. Magn. Magn. Mater.
242-245 (2002) 33.

\bibitem{vons}S. V. Vonsovsky, {\it Magnetism}, Nauka, Moscow, 1971.

\bibitem{boris}B. I. Belevtsev, A. Ya. Kirichenko, N. T. Cherpak,
G. V. Golubnichaya, I. G. Maximchuk, A. B. Beznosov, V. B.
Krasovitsky, P. P. Pal-Val, and I. N. Chukanova,  J. Appl. Phys.
94 (2003) 2459.

\bibitem{hu}Jifan Hu and Hongwei Qin, J. Magn. Magn. Mater. 234 (2001) 419;
Jifan Hu and Hongwei Qin, Mater. Sci. Eng. B 79 (2001) 186.

\bibitem{wang1}Jinhui Wang, Gang Ni, Wenli Gao, Benxi Gu, Xiabin Chen, and
Youwei Du, Phys. Stat. Sol. A 183 (2001) 421.

\bibitem{wang2}Jinhui Wang, Benxi Gu, Hai Sang, and Youwei Du,
J. Magn. Magn. Mater. 223 (2001) 50.

\bibitem{fink}V. Eremenko, S. Gnatchenko, N. Makedonska, Yu.
Shabakaeva, M. Shvedun, V. Sirenko, J. Fink-Finowicki, K. V.
Kamenev, G. Balakrishnan, and D. McK Paul, Fiz. Nizk. Temp. 27
(2001) 1258.

\bibitem{urushibara}A. Urushibara, Y. Morimoto, T. Arima, A.
Asamitsu, G. Kido, and Y. Tokura, Phys. Rev. B 51 (1995) 14103.

\bibitem{mukhin}A. A. Mukhin, V. Yu. Ivanov, V. D. Travkin, S. P.
Lebedev, A. Pimenov, A. Loidl, and A. M. Balbashov, JETP Letters
68 (1998) 356.

\bibitem{boris2}B. I. Belevtsev, Fiz. Nizk. Temp. 30 (2004) 563;
Preprint cond-mat/0308571.

\bibitem{smit}J. Smit and H. P. J. Wijn, {\it Ferrits},
Philips Technical Library, Eindhoven, 1959.

\end{thebibliography}
\end{document}